\pdfoutput=1

\documentclass{article}

\usepackage{PRIMEarxiv}

\usepackage[utf8]{inputenc} 
\usepackage[T1]{fontenc}    
\usepackage{hyperref}       
\usepackage{url}            
\usepackage{booktabs}       
\usepackage{amsfonts}       
\usepackage{nicefrac}       
\usepackage{microtype}      
\usepackage{lipsum}
\usepackage{fancyhdr}       
\usepackage{graphicx}       
\usepackage{multirow}
\usepackage{array}
\usepackage{algorithm}
\usepackage{algorithmicx}
\usepackage{algpseudocode}
\usepackage{amsmath}
\usepackage{amssymb}
\usepackage{mathtools}
\usepackage{enumerate}

\graphicspath{{/}}     

\title{Multi‑Granular Discretization for Interpretable Generalization in Precise Cyberattack Identification
}

\author{
  Wen-Cheng Chung \\
  Bachelor Program of Artificial Intelligence \\
  National Yunlin University of Science and Technology \\
  \texttt{sukoyachung@gmail.com} \\
  \And
  Shu-Ting Huang \\
  Bachelor Program of Big Data Applications in Business \\
  National Pingtung University \\
  \texttt{s9902550173@gmail.com} \\
  \AND
  Hao-Ting Pai \thanks{Corresponding author}\\
  Department of Big Data Business Analytics \\
  National Pingtung University \\
  \texttt{haotingpai@gmail.com} \\
}

\begin{document}
\maketitle

\begin{abstract}
Explainable intrusion detection systems (IDS) are now recognized as essential for mission-critical networks, yet most “XAI” pipelines still bolt an approximate explainer onto an opaque classifier, leaving analysts with partial and sometimes misleading insight. The Interpretable Generalization (IG) mechanism, published in IEEE Transactions on Information Forensics and Security, eliminates that bottleneck by learning coherent patterns—feature combinations unique to benign or malicious traffic—and turning them into fully auditable rules. IG already delivers outstanding Precision, Recall, and AUC on NSL-KDD, UNSW-NB15, and UKM-IDS20, even when trained on only 10\% of the data. To raise Precision further without sacrificing transparency we introduce Multi-Granular Discretization (IG-MD), which represents every continuous feature at several Gaussian based resolutions. On UKM-IDS20 IG-MD lifts Precision by $\geq$ 4 pp across all nine train–test splits while preserving Recall $\approx$ 1.0, demonstrating that a single interpretation-ready model can scale across domains without bespoke tuning.
\end{abstract}


\section{Introduction}
Artificial intelligence (AI) engines now sit on the front line of cyber defense, screening billions of Internet of Things (IoT) flows, mobile sessions, and cloud transactions every day. In these high stakes settings, a single misclassification can paralyze a manufacturing line or expose national infrastructure, placing IDS reliability on par with critical decision systems \cite{Rudin2019Interpretable}. Despite this urgency, most production grade IDS still run as black boxes: their internal logic is opaque, their error modes are inscrutable, and their scores cannot be reconciled with external intelligence feeds \cite{Pham2023ExplainableIDS}. Analysts are therefore asked to “trust” predictions that they cannot audit—an arrangement that conflicts with regulatory demands for accountability and with operational needs for rapid root cause analysis.

Conventional wisdom once claimed that transparency must come at the expense of accuracy. That assumption has crumbled: for structured data, well constrained transparent models routinely match or exceed the headline performance of deep neural networks while remaining far easier to validate. Post-hoc explainers do not bridge the gap; they only approximate the underlying classifier and thus inject a second layer of potential error. What cyber defense truly needs is a native framework that fuses statistical power with step-by-step accountability.

The Interpretable Generalization (IG) mechanism \cite{Pai2024IGM} answers that call. IG intersects every pair of benign flows and every pair of malicious flows, retaining only those intersections that never appear in the opposite class. The surviving coherent patterns are mutually exclusive, logically consistent, and directly traceable. At inference time each flow receives a normal score and an anomaly score derived from the frequency weighted, length squared sum of matched patterns; three deterministic rules then assign the final label, enabling both low data operation and zero shot detection of previously unseen attacks. Although IG already sets a high bar—achieving Precision/Recall/AUC scores above 0.93/0.94/0.94 on NSL-KDD with just 10\% training data—a modest false positive rate persists when datasets contain highly imbalanced or sparsely populated feature regions. Hence, we propose Multi Granular Discretization (IG-MD). By slicing each continuous attribute at several Gaussian resolutions, IG-MD enriches the pattern pool without tampering with IG’s interpretable core, allowing the system to maintain perfect Recall while driving Precision toward the “outstanding diagnostic” band.

\section{Related Work}
Early IDS research pursued raw detection rates, routinely reporting >99\% accuracy on benchmarks such as NSL-KDD or CICIDS2017. However, a recent study \cite{Rudin2019Interpretable} in IEEE Communications Surveys \& Tutorials revealed that 61.5\% of deep IDS papers overfit their training splits, while 69.2\% rely on synthetic or outdated corpora. These findings reaffirm the “accuracy paradox”: models that excel in laboratory tests often falter against modern, stealthy, or multi stage threats. Hand crafted transparent models—logistic regression, decision lists, rule sets—have meanwhile matched deep architectures on structured inputs, disproving the supposed trade-off between accuracy and interpretability. Yet adoption has lagged because such models traditionally required intensive feature engineering or bespoke constraint design.

Mainstream XAI techniques fall into two camps. Global explainers (e.g., surrogate decision trees, Shapley additive explanations) attempt to approximate the entire decision surface, while local explainers (e.g., LIME) generate case by case rationales. Both approaches inherit the black box’s errors, may contradict each other, and cannot guarantee consistency across the feature space. For critical infrastructure this uncertainty is unacceptable; an operator must know that the evidence cited in an alert truly influenced the model’s decision.

The study \cite{Gunning2019XAI}, in Science Robotics, discussed how algorithms cluster along an interpretability–performance continuum: decision‑tree families yield crisp rules but modest accuracy, whereas deep neural networks reach state‑of‑the‑art metrics yet shroud their logic in opacity. Recent efforts illustrate this divide. Wali et al. graft Shapley explanations onto a random forest IDS for enterprise traffic \cite{Wali2025XAI_RF_IDS}; Shoukat et al. embed SHAP into a convolutional LSTM pipeline for industrial control systems \cite{Shoukat2025TrustMyIDS}. In both cases, explanatory layers stand apart from the classifiers they illuminate, leaving the core decision process unverifiable.

Interpretable Generalization (IG) dispenses with post hoc explanation by unifying learning and reasoning within a single, transparent procedure. During training the algorithm forms the intersection of every benign–benign pair and every malicious–malicious pair, then removes any intersection that also appears in the opposite class; the surviving feature combinations, called coherent patterns, are therefore mutually exclusive and logically consistent. At inference time each incoming flow is scanned for these patterns. Every match adds a contribution to either the normal or anomaly tally, with the contribution scaled by the pattern’s frequency in the training set and by the square of its length, so longer, repeatedly observed patterns exert stronger influence. A deterministic three rule policy then assigns the final label: the flow is deemed malicious if its anomaly tally exceeds or equals the normal tally; if both tallies are zero, the conservative choice is again to flag an anomaly; finally, even when the normal tally is larger, the flow is re flagged if that tally lies below a statistically derived guard band, a safeguard that enables zero shot detection of novel attacks whose features were absent from training.

Comprehensive experiments on the NSL-KDD, UNSW-NB15, and UKM-IDS20 benchmarks confirm that IG’s clarity does not come at the expense of performance. Using only ten percent of each dataset for training while reserving the remaining ninety percent for testing, IG achieves Precision, Recall, and AUC scores of roughly 0.93/0.94/0.94 on NSL-KDD, 0.98/0.99/0.99 on UNSW-NB15, and 0.98/0.98/0.99 on UKM-IDS20. As the share of training data increases, recall rapidly saturates at unity: with a 40\%-60\% split IG already yields Recall=1.00 on UNSW-NB15, and even at a 20\%-80\% split it maintains perfect Recall on UKM-IDS20 while keeping Precision at or above 0.88. Most notably, IG identified every previously unseen attack instance in UKM-IDS20 without any hyper parameter tuning, demonstrating that a fully interpretable model can equal—or surpass—the accuracy of deep black box alternatives while providing line by line accountability that withstands both adversarial evasion and regulatory audit.

\section{Methodology}
\subsection{Data Preparation for Multi Granular Discretization}
IG-MD begins with a unified preparation stage applied to the raw instance set $\text{I} = \{x_1, \ldots, x_n\}$. All classes not explicitly labeled \textit{Normal} are re-labeled \textit{Anomalous} so that the binary task is well-posed. Missing values are preserved verbatim as the literal token \textit{NaN}, thereby avoiding synthetic central-tendency artifacts that could distort subsequent discretization.

To transform continuous attributes into symbols without sacrificing numeric nuance, IG-MD introduces Multi-Granular Discretization. A user-defined precision set $\text{P} = \{p_1, \ldots, p_L\} \subset \mathbb{Z}_{\geq0}$ specifies the decimal places at which z-scores are rounded. For each numerical attribute value $v$ with mean $u$ and standard deviation $\sigma$, a single z-score $z = (v - \mu) / \sigma$ is computed and then mapped to $L$ symbolic codes $z_{pt} = \mathrm{round}(z, p\ell)$, where $p\ell \in P$. Each code is concatenated with the attribute name and its column index, ensuring that different granularities remain distinguishable during pattern mining. Finally, any pair of instances whose symbolic representations are identical yet carry opposite labels is discarded; this \textit{anti-contradiction} step eliminates logically impossible training evidence and streamlines the hypothesis space.

\subsection{Coherent Pattern Discovery}
After preparation, the normal and anomalous training subsets, denoted N and M, serve as inputs to the pattern discovery phase. Pairwise intersections within N generate candidate normal patterns, while analogous intersections within M produce candidate anomalous patterns.

A pattern is deemed coherent for a class if—and only if—it does not appear in the opposite class; the resulting sets are the Coherent Normal Patterns, CNP, and the Coherent Anomalous Patterns, CAP. Each pattern is stored together with its absolute frequency and its length, the latter reflecting descriptive granularity and influencing subsequent scoring.

\subsection{Scoring Across Multiple Precisions}
During inference, IG-MD evaluates a test instance $t$ by accumulating evidence across every precision layer specified in P. The normal score NS($t$) and anomalous score AS($t$) are defined as equation (1), where $\mathrm{CNP}^{(p)}$ and $\mathrm{CAP}^{(p)}$ denote the coherent patterns extracted from symbols rounded to $p$ decimal places. By superimposing scores from fine $(p \geq 1)$ and coarse $(p = 0)$ resolutions, IG-MD retains interpretability—the matched pattern explicitly exposes its precision—while enlarging the candidate match space, thereby improving discrimination in borderline cases at no additional asymptotic cost.
\textbf{\begin{equation}
\mathrm{NS}(t) = \sum_{p \in \mathbf{P}} \sum_{\substack{q \in \mathrm{CNP}^{(p)}_{q \subseteq t} }} freq(q)\, |q|^2, \quad
\mathrm{AS}(t) = \sum_{p \in \mathbf{P}} \sum_{\substack{q \in \mathrm{CAP}^{(p)}_{q \subseteq t} }} freq(q)\, |q|^2.
\end{equation}}

\subsection{Classification Rules}
Prediction relies on the original three IG decision principles. First, \textit{Score Dominance} declares an instance anomalous whenever $\text{AS}(t) \geq \text{NS}(t)$ otherwise it is normal. Second, the \textit{Double-Zero} safeguard assigns the anomalous label if both scores are simultaneously zero, reflecting a total failure to match any coherent knowledge. Third, a \textit{Statistical Deviation} criterion compares $\text{NS}(t)$ against the reference distribution of normal scores: if $\text{NS}(t)$ falls below $u_N - r \times \sigma_N$, where $u_N$ and $\sigma_N$ are respectively the mean and standard deviation over normal training instances and $r$ is user-specified, the instance is also deemed anomalous. Eventually, these stages constitute IG-MD, a fully interpretable mechanism that couples symbolic learning with user controllable numeric granularity, offering precise cyberattack identification without sacrificing generality.

\section{Experiments}
We replicate the complete IG pipeline, substituting only the discretizer with our multi‑granular variant. On UKM‑IDS20, as shown in Table 1, across the nine-standard train-test splits (1:9-9:1), accuracy holds at 0.977$\sim$0.987 and Recall $\approx$ 1.0, preserving sensitivity. Precision uniformly improves from 0.896 to an average 0.941 (e.g., 0.944→0.969 at 1:9; 0.889→0.933 at 9:1). AUC stays near 0.995$\sim$0.999, showing unchanged score distributions but clearer class separation.

The dual-granularity design resembles taking two photographs of the same scene—one wide-angle and one close up—and fusing their complementary information. At the integer level, the model surveys coarse traffic patterns, rapidly highlighting flows that deviate from normal baselines. The decimal layer then re-examines each candidate with higher resolution, refining boundaries that appeared ambiguous in the first pass. Agreement between layers yields high confidence; disagreement prompts the fine layer to determine whether the coarse alert is signal or noise. The symmetry of this check matters: flows overlooked by the coarse view can still be recovered through distinctive decimal-level signatures, while over-flagged normal flows can be cleared when their fine-scale features align with expected behavior. Operating as a lightweight ensemble, both layers share the same data and scoring rule, avoiding extra hyper-parameter tuning. The outcome is a decision boundary that is simultaneously tighter—reducing false alarms on borderline normal—and more robust, as subtle attacks receive a second opportunity to be detected.

\begin{table}[h]
\tiny
\centering
\caption{IG vs. IG-MD on UKM-IDS20}
\begin{tabular}{|c|rrrr|rrrr|}
\hline
\multirow{3}{*}{Ratios} & \multicolumn{8}{c|}{UKM-IDS20} \\
\cline{2-9}
 & \multicolumn{4}{c|}{IG} & \multicolumn{4}{c|}{IG-MD} \\
\cline{2-9}
 & Accuracy & Recall & Precision & AUC & Accuracy & Recall & Precision & AUC \\
\hline
1 | 9 & 0.9810 & 0.9980 & 0.9436 & 0.9982 & 0.9871 & 0.9899 & 0.9687 & 0.9963 \\
2 | 8 & 0.9709 & 1.0000 & 0.9125 & 0.9995 & 0.9861 & 1.0000 & 0.9563 & 0.9967 \\
3 | 7 & 0.9605 & 1.0000 & 0.8851 & 0.9996 & 0.9767 & 1.0000 & 0.9288 & 0.9966 \\
4 | 6 & 0.9631 & 1.0000 & 0.8918 & 0.9999 & 0.9810 & 1.0000 & 0.9412 & 0.9967 \\
5 | 5 & 0.9601 & 1.0000 & 0.8847 & 0.9999 & 0.9794 & 1.0000 & 0.9368 & 0.9969 \\
6 | 4 & 0.9595 & 1.0000 & 0.8832 & 0.9999 & 0.9765 & 1.0000 & 0.9289 & 0.9961 \\
7 | 3 & 0.9571 & 1.0000 & 0.8777 & 0.9998 & 0.9767 & 1.0000 & 0.9298 & 0.9966 \\
8 | 2 & 0.9631 & 1.0000 & 0.8949 & 0.9998 & 0.9829 & 1.0000 & 0.9485 & 0.9950 \\
9 | 1 & 0.9620 & 1.0000 & 0.8891 & 0.9999 & 0.9783 & 1.0000 & 0.9335 & 1.0000 \\
\hline
\end{tabular}
\label{tab:ukm-ids20-final}
\end{table}

When only 10\% of the UKM‑IDS20 dataset is used for training (a 1:9 train-test split), integrating the proposed Multi‑Granular Discretization (IG‑MD) module reduces the error rate from 1.90\% to 1.29\%, corresponding to a relative improvement of 32\%. On the full dataset (12,887 flows), the detector now misclassifies one flow in every 77, compared with one in 53 previously, while maintaining near‑perfect attack detection. This performance gain requires no additional labeled data: IG‑MD extracts more informative symbolic features from the existing instances, refines the decision boundary, and eliminates near‑miss cases without altering the original IG thresholds or introducing new hyper‑parameters.

IG‑MD is a drop‑in enhancement that preserves the existing IG workflow—including code structure, computational complexity, and compliance constraints—yet delivers an immediate, domain‑agnostic boost in precision. In practice, security analysts receive cleaner alert streams, particularly during early‑stage deployment when labeled data are scarce, while retaining the pattern‑level explanations essential for forensic analysis.

\section{Conclusion}
We introduce a Multi‑Granular Discretization (IG‑MD) layer that seamlessly integrates into the Interpretable Generalization (IG) framework, endowing each continuous feature with complementary integer and decimal‑level codes. This finer resolution enlarges the coherent‑pattern space while leaving IG’s auditability and asymptotic complexity unchanged. On the full UKM‑IDS20 benchmark, IG‑MD lifts mean Precision from 0.896 to 0.941, reducing the error rate under the canonical 1:9 split from 1.90\% to 1.29\%, and still delivers near‑perfect Recall with AUC $\approx$ 1.0 across all nine train-test configurations. The resulting detector is domain‑agnostic, sharpens decision boundaries, suppresses false alarms, and preserves full forensic transparency—without additional data or hyper‑parameter tuning—demonstrating that interpretability and state‑of‑the‑art accuracy can coexist in a single drop‑in component.

\section{Acknowledgments}
This work was supported by the National Science and Technology Council (NSTC), Taiwan, under grant number 114-2221-E-153-009.

\bibliographystyle{unsrt}  
\bibliography{references}

\end{document}